\documentclass[10pt,compsoc]{IEEEtran}
\usepackage[labelfont=bf,textfont={bf}]{caption}
\usepackage{textcomp}
\usepackage{dblfloatfix}  
\usepackage{cite}
\usepackage{algorithmic}
\usepackage[table]{xcolor}
\usepackage{graphicx}  
\usepackage{balance}
\usepackage{wrapfig}
\usepackage{url}
\usepackage[most]{tcolorbox}
\newtcolorbox{hence}[1][]{%
  colback=gray!10,        
  colframe=black,         
  left=2mm,               
  boxrule=0mm,            
  frame hidden,           
  borderline west={1mm}{0mm}{black}, 
  width=\linewidth,       
  enhanced,               
  #1                      
}

\usepackage{enumitem}
\setitemize{noitemsep,topsep=0pt,parsep=0pt,partopsep=0pt}
\newcommand{\bi}{\begin{itemize}}
\newcommand{\ei}{\end{itemize}}

\usepackage{mdframed} 
\usepackage{fontawesome5}  
\usepackage{amssymb}       

\definecolor{grayline}{gray}{0.8}

\begin{document}
 
\title{Retrospective: Data Mining Static Code Attributes to Learn Defect Predictors}

\author{ Tim~Menzies,~\IEEEmembership{Fellow,~IEEE}
\IEEEcompsocitemizethanks{\IEEEcompsocthanksitem   T. Menzies is with the Department
of Computer Science, North Carolina State University, Raleigh, USA.
 \protect
E-mail: timm@ieee.org}}

\IEEEtitleabstractindextext{
\begin{abstract}
Industry can get any research it wants, just by publishing a baseline result along with the   data and scripts need to reproduce that work. For instance, the   paper  ``Data Mining Static Code Attributes to Learn Defect Predictors'' presented such a baseline, using static code attributes from NASA projects. Those result were enthusiastically embraced by a software engineering research community, hungry for data. At its peak (2016) this paper was SE's most cited paper (per month).  By 2018,   twenty percent of  leading TSE papers (according to Google Scholar Metrics), incorporated artifacts introduced and disseminated by this research. This brief note reflects on what we should remember, and  what we should forget, from that paper.
\end{abstract}

}

\maketitle 
\IEEEpeerreviewmaketitle
\IEEEdisplaynontitleabstractindextext

\section{Introduction}

\begin{wrapfigure}{r}{0.65in}
\noindent\includegraphics[width=0.65in]{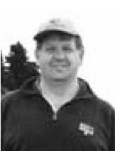}\vspace{3px}
\includegraphics[width=0.65in]{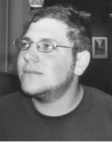}\vspace{2px}
\includegraphics[width=0.65in]{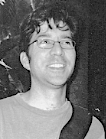}
\end{wrapfigure}
The 2007   TSE paper    ``Data Mining Static Code
Attributes to Learn Defect Predictors''~\cite{Menzies07} 
by Tim
Menzies,  Jeremy Greenwald   and Art
Frank was born of the open source culture at Portland, Oregon.  
In that halycon  time, our clothes were   damp after   
push biking in the rain to coffee shops for Ruby-on-Rails meet-ups.  We wore no suite and tie in our photos.  
We did not comb our hair. Instead,  we wannabe  larrikins stood barefoot on the   beach,    with   bike messenger bags over our \mbox{t-shirts},
united in the belief that   
\begin{quote}
{\tt  svn commit -m "share stuff"}
\end{quote}
will bring down the evil empire, one GNU public license at a time.

One night in 2004,  walking  around Chicago's Grant Park, 
the open source culture meet SE research.
Jelber Sayyad  and I were
lamenting the sad state of machine learning in SE.   ``Must do better'', we said.  ``Why don't we  make conclusions reproducible? Make authors promise to    publish their   data with their papers?"

In 2025 it is  hard to believe that this idea of ``reproducible SE'' was a radical idea.  But at that time, there was little sharing -- so much so that in 2006 Lionel Briand predicted it will not work, famously saying ``no one will give you data''.

Nevertheless, perhaps influenced by the emerging power of the open source economy,
Jelber and I persisted and created the
  PROMISE project. It had  two parts:
\bi
\item An annual conference on  predictor models in software engineering  (to share results and study open issues)\footnote{\url{https://conf.researchr.org/series/promise}}.
\item A repository of 100s of SE data sets about defect prediction, effort estimation,
Github issue close time, bad smell detection and dozens of other topics\footnote{http://tiny.cc/promise25}. 
This repository grew   so large and that   it we   moved it to the Large Hadron Collider (see the ``Seacraft'' data at  Zenodo\footnote{\url{https://zenodo.org/communities/seacraft/records?q=&l=list&p=1&s=10&sort=newest}.}). These days it is somewhat inactive but in its heyday,
my research students  ran regular  week-long sprints that scoured the table of content of recent SE conferences and journals, reaching out to authors for    their reproduction data\footnote{\url{https://github.com/opensciences/opensciences.github.io/issues}}. 
\ei
At first, the PROMISE series got off to a shaky  start. But once Gary  Boetticher,
Elaine Weyuker, Thomas Ostrand,  and Guenther Ruhe~\cite{MENZIES2024107486}
joined the steering committee, the meeting earned the prestige needed for future  growth.

In those early days, it
was very encouraging to see so many researchers
taking up the idea of reproducible results.
 While other research areas struggled to obtain reproducible results, PROMISE swam (as it were) in an ocean of reproducibility.
Numerous papers were written that applied an increasing
elaborate tool set to data like COC81, JM1, XALAN, DESHARNIS
and all the other data sets that were used (and reused) in the first decade of PROMISE.

Concurrently with PROMISE's development, was MSR, a similar initiative devoted to Mining Software Repositories\footnote{https://conf.researchr.org/series/msr}~\cite{Hassan2008TheRA}. According to Devanbu~\cite{DEVANBU2015vii}, the MSR conference was primarily occupied with gathering initial datasets from software projects. In contrast, the focus of the PROMISE community was placed on the post-collection analysis of this data. Typically, MSR publications did not prioritize revisiting datasets already analyzed in previous work~\cite{robles2010replicating} (a trend that has improved, only somewhat~\cite{GONZALEZBARAHONA23}). Conversely, PROMISE contributors consistently uploaded all their data to a public repository and their subsequent publications often re-examined existing data to refine the analysis.

This 
   PROMISE-style of research  lead to many successful papers. 
   Our 2007   TSE publication was one such paper. By  2018, 20\% of the articles listed in {\em Google Scholar Software Metrics} for  IEEE Transactions
on SE used data sets from that first decade of PROMISE.

\section{What did the 2007  Paper Say?}
That paper explored   data mining algorithms to learn software defect predictors from static code attributes. Why do that?
Well,
data miners can input
features extracted from source code and output predictors for where defects are likely to occur.
While such predictors are never 100\% correct,
they can suggest where to focus on more expensive methods. 
This is useful since software quality assurance budgets are finite while assessment effectiveness increases exponentially with   effort~\cite{fu2016tuning}. Standard practice is to apply slower methods on code sections that seem most critical or bug-prone.
Software bugs are not evenly distributed across a project~\cite{hamill2009common,koru2009investigation, ostrand2004bugs,misirli2011ai}. Hence, 
a useful way to test software   is to allocate most assessment budgets to
the more defect-prone parts of the code (as indicated by defect predictors).

To better understand   defect prediction, the 
 paper   offered  counter arguments to  two prominent prevailing views:
\bi
\item  {\em Specific metrics matter. }
\item   {\em Static code attributes do not matter. }
\ei
In the 1990s, before researchers  had access to extensive SE data, there were prolonged, somewhat heated, theoretical debates on the value of metric~$X$ vs metric~$Y$ (e.g.~\cite{Fenton94}).
So 
to test if (e.g.) McCabe's cyclomatic complexity metrics~\cite{mccabe1976complexity} were any better than (e.g.) Halstead readability metrics~\cite{halstead1977elements}, our 2007  paper applied feature pruning. That is, if an attribute did not improve model performance, it was discarded. For a set of   three dozen metrics, and seven data sets,  pruning selected just  two or three attributes. 
In the selected sets, there was no evidence that (e.g.) Halstead was better than (e.g.) lines of code measures since different data sets selected for different attributes (and no single attribute was selected in the majority of data sets).
Hence:
 \begin{hence}
{\bf Menzies's 1st Law:}  Specific metric \underline{\bf do not} always mattered in all data sets. Rather,   different projects have different best metrics.  
 \end{hence}
This leads to the following
process recommendation:
  \begin{hence}
{\bf Menzies's Corollary:}  To mine SE data, gather   all can that be collected (cheaply) then apply data pruning to discard irrelevancies.   
 \end{hence}
 As to other work,
 Fenton and Pfleeger had  examples of the same functionality   achieved via different  language constructs resulting in different static measurements~\cite{fenton1997metrics}. They used these examples to argue the uselessness of static code attributes. Similarly, Sheppard and Ince~\cite{shepperd1994critique} had   correlation results showing that 
``for a large class of software   (static code measures) are no more than a proxy for, and in many cases outperformed by, lines of code.''

In stress test these views, our 2007 paper 
   first documented current baselines in defect prediction.
   Then, it went on to  show that detectors learned from  static code attributes (using public domain data miners) did much better than those baselines.
   Further, models built from more one attribute did better than single-attribute models. Hence:
   
   \begin{hence}
 {\bf Menzies's 2nd Law:}  Static code attributes \underline{\bf do} matter. Individually, they may be     weak indicators. But when combined, they can lead to a strong  signals
that outperform the state-of-the-art.
\end{hence}

(Aside: lately it has became clear that while
different kinds of code attributes do not matter, one class of ``process-level'' metrics
might matter more~\cite{Majumder22,rahman2014comparing}.)

Another  contribution of the 2007 paper was  methodological.
It defined a set of   steps to build and report the results of data mining. Then its
conclusion
  begged the research community to  try and out-perform its results: 
\begin{quote}
{\em ``Paradoxically, this paper will be a success if it is quickly superseded.''}
\end{quote}
To support that, the paper shared all its   scripts and data. 
As such, it became a handy ``go away and try this!'' document that a hundred supervisors could give to a thousand graduate students.
This perhaps explains the popularity of this paper: at its peak in 2016, this
work was the most cited (per month) paper in software engineering. At
the time of this writing, that 2007 paper~\cite{Menzies07} and the PROMISE  repository\footnote{http://tiny.cc/promise25} have 1924 and 1242 citations (respectively) in Google Scholar.

%


\section{Progress Since 2007}

Since that paper, interest in defect prediction has only increased.
In their 2018 survey of 395 commercial practitioners from 33 countries and five continents,
Wan et al.~\cite{wan18} found that over 90\% of the respondents were willing to adopt defect prediction techniques. 

 Results from commercial projects have shown the benefits of defect prediction.
Misirli et al.~\cite{misirli2011ai} built a defect prediction model for a telecommunications company. Their models predicted 87\% of code defects and decreased inspection efforts by 72\% (while reducing post-release defects by 44\%). Kim et al.~\cite{kim2015remi} applied the defect prediction model, REMI, to the API development process at Samsung Electronics.
Their models could
predict the bug-prone APIs with reasonable accuracy~(0.68 F1 scores) and reduce the resources required for executing test cases.

Software defect predictors not only save labor compared with traditional manual methods, but they are also competitive with certain automatic methods. 
Rahman et al. ~\cite{rahman2014comparing} compared (a) static code analysis tools FindBugs, Jlint, and PMD with (b) defect predictors (which they called ``statistical defect prediction'') built using logistic regression.
No significant differences in cost-effectiveness were observed.

Given this equivalence, it is significant to
note that defect prediction can be quickly adapted to new languages by building lightweight parses to extract code metrics. The same is not true for static code analyzers - these need extensive modification before they can be used in new languages.

Because of this ease of use, and its wide applicability, defect prediction has been  extended many ways:
\bi
\item Application of defect prediction methods to locating code with security vulnerabilities~\cite{Shin2013}.
  \item Predict the location of defects so that appropriate resources may be allocated (e.g.~\cite{bird09reliabity})
  \item Understand the factors that lead to a greater likelihood of defects such as defect prone software components using code metrics (e.g., ratio comment to code, cyclomatic complexity) \cite{menzies10dp, menzies07dp, ambros10extensive} or process metrics (e.g., number of changes, recent activity) \cite{nagappan05codechurn,elbaum00codechurn, moser08changemetrics, hassan09codechanges}. 
  \item Use  predictors to proactively fix defects~\cite{kamei16_lit, legoues12_aprlit, arcuri2011practical}
  \item Study defect prediction not only just release-level \cite{di18_fft, agrawal2018better} but also change-level or just-in-time \cite{yan18_tddetermination, kamei12_jit, nayrolles18_clever, commitguru} both for research and also industry. 
  \item Explore ``transfer learning'' where predictors from one project are applied to another~\cite{krishnaTSE18,nam18tse}.
  \item Explore the trade-offs between explanation and performance of defect prediction models~\cite{di18_fft}.
  \item Assess different learning methods for building models that predict software defects~\cite{ghotra15}. This has led to the development of hyperparameter optimization and better data harvesting tools \cite{agrawal2018wrong, agrawal2018better, Fu17easy, Fu16Grid, fu2016tuning,tantithamthavorn2016automated}. 
\ei

\section{From Success to Stagnation}
The success of the 2007 paper had an unwanted side-effect.
It turns out that when something gets very successful, it tends to get copied ad nauseam.
 So it is no surprise that like many repositories of reproducible case studies, the PROMISE data went through  four phases: 
\bi
\item {\bf ``Data? Good luck with that!''} – Early attempts to share data are met with resistance, skepticism, or outright refusal (e.g., see   Briand's comment in the introduction).
\item {\bf ``Okay, maybe it's not completely useless.''} – The value of the data is grudgingly acknowledged.
\item {\bf ``This is the gold standard now.''} – The data is a required baseline, dictating   the norms of a field.
\item {\bf ``A graveyard of progress.''} – What was once a lifeline is now a lead weight, stifling creativity, and  locking researchers into outdated paradigms.
\ei
Sadly,       decade two of PROMISE, many researchers   continued that kind of first-decade research. Too often, I must review papers from authors who just
  use (e.g.) the COC81 data set   published in 1981~\cite{boehm81}; the DESHARNIS data set,  first published in 1988~\cite{desharnais1988statistical}; the JM1 data, first published in 2004~\cite{menzies2004good}; or the XALAN data set, first published in 2010~\cite{Jureczko10}.
 
 Just to be clear,
there is value in   a publicly accessible set of reference problems. For instance, if a PROMISE author is unable to present results from confidential industrial data, they can use the reference collection to construct a reproducible example of their technique.

That said, there can be too much use 
of a shared resource.
We need to move on from  on decades-old PROMISE data.
In 2025, we have access to much more recent information\footnote{
E.g. see the 1100+ recent Github projects
used by Xia et al.~\cite{xia22},
or everything that can be extracted using CommitGuru\cite{rosen2015commit}.}.
Accordingly, recently I changed the editorial policy at the Automated Software Engineering journal:
we now desk reject  papers based on   datasets   I collected in    2005\footnote{CM1, JM1, KC1, KC2, KC3, KC4, MC1, MC2, MW1, PC1, PC2, PC3, PC4 and PC5.}.

\section{Future Work}

Several steps are being taken to address the  problems with PROMISE.
The annual PROMISE meeting knows it needs to   revisit its
goals and methods. 
Gema Rodr\'iguez-P\'erez\footnote{Member, PROMISE steering committee} notes that, in 2025,
data sharing and replication packages are expected for almost
all SE papers. This means there is now  little distinction between PROMISE and other conferences. 
So, she says,  PROMISE must reverse the trend where
new papers do not offer new data.
While current datasets do offer value, PROMISE should look to accepting higher quality datasets than typical conferences. Perhaps PROMISE authors can consider enhancing their current data space or   conducting more evaluations on its quality.  

That said, Steffen Herbold\footnote{Member, PROMISE steering committee. PROMISE'23 PC co-chair. } cautions that in the early years of PROMISE, data sets were often not really raw data, but rather directly collections of metrics. In MSR, this shifted: data sets like GHtorrent were rather raw data and augmented with fast tools (e.g., PyDriller). These means that more and more researchers and moving towards on-the-fly data collection, further reducing the need for data sharing. The drawback here is obvious: little curation, little validation, often purely heuristic data collection without quality checks even in case of known problems. Thus, he warns, all this newer data might not necessarily be better~\cite{RODRIGUEZPEREZ2018164,herbold2022problems}.  
 
\subsection{What's New and Hot }

Current results suggest  exciting research directions.

For a ``fast forward'' to see how  contemporary researchers addresses
the same problem as the 2007 paper,
see ``DeepLineDP: Towards a Deep Learning Approach for Line-Level Defect Prediction'' 
by Pornprasit et al. ~\cite{Pornprasit23}
(this was a  TSE best paper award winner for 2023). 

Model interpretability is another significant challenge in the field.   It is encouraging to see that more research is being conducted to address this issue~\cite{tantithamthavorn2021explainable}.

Further,  at its core, defect prediction as described in 2007 was a binary classification problem. 
But 
software engineering tasks rarely involve a single goal.
Hence, since that paper,  I spend less time in classification than in  multi-objective optimization for  hyperparameter selection~\cite{xia22}  
or unfairness reduction~\cite{chakraborty2020fairway,Alvarez23}
or determining good management decisions for a software process~\cite{RE-2002-FeatherM,Menzies09}. That research eschews classifiers and, instead, uses  CPU-intensive algorithms like    MaxWalkSat~\cite{Menzies09}, 
 simulated annealing~\cite{Menzies07a,RE-2002-FeatherM} or genetic algorithms.

Furthermore, all the above often assumes
that
analysts can access a large number of good quality labels. Increasingly, I have been growing more and more suspicions of that assumption. These days, my research focus is on how much can be achieved in software engineering using as little data as possible.  This work explores
methods like  landscape analysis~\cite{Chen19,lustosa2024learning},  surrogate learning~\cite{Nair}  active learning~\cite{Krall15,yu18}, 
and semi-supervised learning~\cite{Tu22,Majumder22}.

\subsection{Stranger Things}

\begin{raggedleft}{\em 
In any field, find the strangest thing,and then explore it.} 

--John Archibald Wheeler, physicist.

\end{raggedleft}

Another way to improve future research is to explore anything strange  seen in past results. And there is a long list of strange results from PROMISE.

For example,
consider transfer learning research~\cite{turhan2009relative}
where 
   models from Turkish white goods were successful at predicting errors in   NASA   systems. 
 Transfer learning  is often seen as complex multi-dimensional transform that maps attributes in one domain to another~\cite{pan2009survey}.  But for defect prediction, all that was needed was some simple nearest neighboring between   test data     and training data and 
voil\`a:
 \begin{hence}
{\bf Menzies's 3rd Law:} Turkish toasters can  predict for errors in deep space satellites. 
 \end{hence}
Perhaps the lessons here is that  many of the distinctions made about software are spurious and need to be revisited.

Another  strangeness, seen in the 2007 paper. as well as subsequent work~\cite{menzies2021shockingly,rees2017better,Kocaguneli13aa,7194627,chen2005finding},
  was that  pruning rows and columns results in better models.
Readers  familiar with the manifold assumption~\cite{zhu2005semi} and the Johnson-Lindenstrauss lemma~\cite{johnson1984extensions} will be nodding sagely at this point-- but the reductions  seen in SE data are   startling. 
For example, 
Chen, Kocaguneli,   Tu,   Peters, and Xu et al. 
found they could
predict for
Github issue close time,
effort estimation, and 
defect prediction,  even after ignoring      labels  for 80\%, 91\%, 97\%, 98\%, 100\% (respectively) of their  project data labels~\cite{chen2005finding, Kocaguneli13aa,Tu22,7194627,XU2021110862}.
Data sets with thousands of rows can be modeled with just a few dozen samples~\cite{menzies2008implications}-- perhaps because of power laws~\cite{lin2015power} or large amounts of repeated structures~\cite{hindle} in the data from SE projects.
So we really need to study why:
 \begin{hence}
{\bf Menzies's 4th Law:}  For SE, the best thing to do with most data is  to  throw    it     away.
 \end{hence}
Of course,  here I am talking about regression~\cite{chen2005finding}, classification~\cite{rees2017better} and optimization~\cite{chen18}.
Generative tasks may require models with billions of variables learned from 100s of gigabytes of data.
But while I am talking about LLMs: 
 \begin{hence}
{\bf Menzies's 5th Law:}  Bigger is not necessarily better.
 \end{hence}
There is  much LLM hype these days but very little comparison of LLMs to other methods.  
For example, in  a recent  systematic review~\cite{Hou24} of 229 SE papers using large language models (LLMs),  only $13/229 \approx 5\%$ of those papers compared LLMs to other approaches. This is a methodological error
since   other methods (developed and certified using PROMISE-style research) can
 produce results that are better and/or 
faster~\cite{grinsztajn2022why,somvanshi2024survey,Tawosi23,majumder2018500+,Ling24,Fu17,johnson2024ai}.

Next, there is  data quality. 
For PROMISE data:
\begin{hence}
{\bf Menzies's 6th Law:} Data  quality matters less than you think.
 \end{hence}
 Data collection   must be done with  care. But there is such a thing as too much care.
 Effective predictions can be made from seemingly dirty data. In 2013, Shepperd et al.~\cite{shepperd2013data} found   numerous quality issues with PROMISE data (e.g. repeated rows, illegal attributes,etc.). But they   never tested  if 
{\em increasing} their kinds of quality issues
{\em decreased} the predictive power of learned models. To address that, we built mutators that injected an increasing amount of   their quality issues into PROMISE defect data sets. 
Strangely, the performance curve remained flat despite the increased number of quality issues.  Which is really strange.

A related strangeness  is this:
\begin{hence}
{\bf Menzies's 7th Law:} Bad learners can   make good conclusions.
 \end{hence}
When exploring CART trees built to guide multi-objective optimization, Nair et al.~\cite{nair2017using} found that models that predicted poorly   could still   rank one solution over another. Hence, they can be used to (e.g.) prune away poor configurations in order to find better ones. This  suggests that the algorithms we are using to explore data are missing the point. Maybe they should not be aiming to make predictions but instead offer   weak hints about project data?

Moving along:
\begin{hence}
{\bf Menzies's 8th Law:}  Science has mud on the lens.
 \end{hence}
 One of the lessons of hyperparameter optimization~\cite{agrawal2018better,agrawal2019dodge,tanti16defect,Fu16aa} on PROMISE data is that   conclusions reached via data mining can be changed and made more accurate, in an afternoon, by a grad student with enough CPU. Does this mean all our conclusions are brittle and prone to reversal at any time?
  How can we build a scientific community on such a basis? 
 Where are the stable conclusions that can be used to build tomorrow's ideas? Our Bayesian colleagues may  have much to say on this topic.

 Finally here is the  strangest thing I have seen in all my years working at this kind of data:
\begin{hence}
{\bf Menzies's 9th Law:}  Many hard SE problems, aren't.
 \end{hence}
 In his book  {\em Empirical  Methods~}\cite{cohen1995empirical}, Cohen argues that supposedly sophisticated methods should be    benchmarked against seemingly stupider ones (the so-called ``straw man'' approach to scientific verification).  I can attest that  whenever I checked a supposedly sophisticated method against a simpler one, there was always something useful in the simpler. And more often than not,  a year later, I have switched to the simpler approach~\cite{agrawal2018better,agrawal2019dodge,tanti16defect,Fu16aa}. 
 
The caveat here is that not all SE problems can be simplified. For example,
  generation tools  probably   need the complexities of LLMs. Also,
the certification requirements of safety-critical
software is not a simple process.
But just because some tasks are hard,
does not mean all tasks are hard.   So I challenge  the research community: 
\begin{quote}
{\em Have we really checked what   is really complex and what is really very  simple?}
\end{quote}

\section{Conclusion}

There is much remaining to learn  from that 2007 paper. 
The spirit of the old Portland 
open source community can still guide us. 
  Many important insights are  obtainable from    PROMISE-style ``do-it-then-do-it-again'' research.
Open science  communities can be formed to  explore any  research topic, just by publishing a baseline result plus the data and scripts needed to reproduce
that result.
And looking at the above list
of strange things,  
 we can see that there  is much left to explore.

\section*{Acknowledgment}

I gratefully acknowledge the hard
work of Jeremy Greenwald and Art Frank,
my co-authors on the 2007 paper.
Thanks also to Gema Rodr\'iguez-P\'erez and Steffen Herbold for their
thoughts on the current state, and future, of 
the PROMISE conference.
And special thanks to all  my research students who worked for so many decades on the   PROMISE project 
(sadly, too many to list here).

\bibliographystyle{IEEEtran}
\bibliography{main}

\begin{IEEEbiography}[{\includegraphics[width=1in,clip,keepaspectratio]{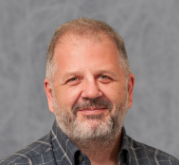}}]{Tim Menzies} (ACM Fellow, IEEE Fellow, ASE Fellow, Ph.D., UNSW, 1995) is a full Professor in Computer Science at North Carolina State. He is the director of the Irrational Research lab (mad scientists r'us) and the author of over 300 publications (refereed) with 24,000 citations and an h-index of 74. He has graduated 22 Ph.D. students, and has been a lead researcher on projects for NSF, NIJ, DoD, NASA, USDA  and private companoes (total funding of \$19+ million). Prof. Menzies is the editor-in-chief of the Automated Software Engineering journal and associate editor of TSE and other leading SE journals. For more, see  \url{https://timm.fyi}.
\end{IEEEbiography}

\end{document}